# A Physics-Informed Data-Driven Fault Location Method for Transmission Lines Using Single-Ended Measurements with Field Data Validation

Yiqi Xing, *Student Member, IEEE,* Yu Liu*, *Senior Member, IEEE,* Dayou Lu, Xinchen Zou and Xuming He

*Abstract*—Data driven transmission line fault location methods have the potential to more accurately locate faults by extracting fault information from available data. However, most of the data driven fault location methods in the literature are not validated by field data for the following reasons. On one hand, the available field data during faults are very limited for one specific transmission line, and using field data for training is close to impossible. On the other hand, if simulation data are utilized for training, the mismatch between the simulation system and the practical system will cause fault location errors. To this end, this paper proposes a physics-informed data-driven fault location method. The data from a practical fault event are first analyzed to extract the ranges of system and fault parameters such as equivalent source impedances, loading conditions, fault inception angles (FIA) and fault resistances. Afterwards, the simulation system is constructed with the ranges of parameters, to generate data for training. This procedure merges the gap between simulation and practical power systems, and at the same time considers the uncertainty of system and fault parameters in practice. The proposed data-driven method does not require system parameters, only requires instantaneous voltage and current measurements at the local terminal, with a low sampling rate of several kHz and a short fault time window of half a cycle before and after the fault occurs. Numerical experiments and field data experiments clearly validate the advantages of the proposed method over existing data driven methods.

## I. INTRODUCTION

Transmission line fault location is critical for fast restoration of power system after the occurrence of the fault, to ensure power supply reliability. Transmission line fault location methods can be mainly divided into traveling wave-based methods, model-based methods, and data-driven methods.

### A) Traveling Wave-Based Methods

The traveling wave based method determines the fault location by detecting the time when the wavefront arrives at the line terminal, including single-ended [1-2] and dual-ended [3-4] methods. The detection reliability could be compromised during zero inception angles or when the fault resistance is high. Also, these methods require very high sampling rate (typically in the order of MHz) for accurate fault location, which may not be available in practice.

### B) Model-Based Methods

Model-based methods find the fault location by establishing transmission line models and solving for the fault location as an unknown. Fundamental frequency phasor-based methods are widely adoptd in practical power systems, including single-ended methods (such as simple reactance method, Takagi method, Eriksson method, etc. ) [5-6] and double-ended [7-8] methods. The accuracy of model based methods is usually dependent on accurate mode      and faults. In addition, phasor domain methods assume sinusoidal steady state operation of the system and need to accurately extract phasors. Therefore, they will experience increased errors when the available time window during fault is short, i.e., the line is equipped with the fast-tripping relays.

### C) Data-Driven Methods

Data-driven methods, also known as artificial intelligence-based methods, have attracted increasing attention in recent years in the area of fault location. Those methods typically adopt a large amount of data or a large-scale dataset to train a useful network. Basically, the data-driven methods can work with a relatively low sampling rate compared to the traveling wave-based methods, and could potentially show robustness against model uncertainties compare to the model-based methods. One of the early attempts of applying the data-driven methods to fault location is achieved in [14], where the input of the network is the raw data of the line terminal current measurements. Reference [15] proposed a neural network (NN) based transmission line fault location method with single-ended raw data. Reference Literature [16] proposed a graph neural network (GNN) based transmission line fault location method combined with a sequence model and data graph structure. Reference [17] uses a Decision Tree Regression (DTR) method to locate the transmission line fault. Discrete wavelet transform and discrete Fourier transforms were used to process the signal. Reference [18] proposed a transfer function based fault location method using convolution neural network (CNN) and deep reinforcement learning (DRL). Reference [19] proposed a high-voltage DC fault location method based on 1D-CNN which uses one-dimension features after Empirical Mode Decomposition (EMD) for the training set. Reference [20] proposed a location method using the double-ended measurements with dynamic state estimation (DSE).

In fact, all the above data driven methods utilize simulation data for training and testing: the training set is simulation data, and the testing set is also from the same simulation system. Specifically, existing literature rarely presents data driven transmission line fault location approaches that are validated using field data, and there is still a huge gap for those approach to be applied to fault location in practical power systems. The reasons are as follows. *On one hand*, the number of faults for a specific transmission line is very limited, making it close to impossible to obtain a well-trained network using only field data. *On the other hand*, if simulation data are utilized for training, since the learned feature space   is usually different from that of the practical data, accurate fault location could not be achieved when applied to practical power systems.

*D) Proposed Method*

This paper proposes a physics-informed data-driven fault location method that is specifically designed for practical power systems and is validated using the field data during line faults. Only one set of field data with three phase voltags and currents measured at the local terminal is required. The required time window is only half a cyle before and after the fault occurs. *First*, with the field data, the proposed method estimates the ranges of system parameters, such as the ranges of source impedances, loading conditions, fault inception angles (FIAs) and fault resistance, through physical models of the line. *Second*, extensive simulated data during faults are generated using power system electromagnetic transient (EMT) simulation tools, with the ranges of system parameters estimated in the first step. Those data formulates the "target dataset" with features consistent with the field data. In this case, the mismatch between the simulation and the practical power system is minimized. *Third*, a deep neural network (DNN) is adopted as an example and the "target dataset" is applied for training to achieve fault location. Simulation results and field data experiments validate improved reliability and the accuracy of the proposed method in comparison to the existing approach. The primary contributions of this paper includes,
1) The inherent physical information of the system is extracted to estimate the ranges of system and fault parameters of the field data; the extraction only requires single-ended measurements within half a cycle before and after the fault occurs, and does not require actual fault location of the system;
2) The "target dataset" is generated via simulation using the estimated ranges of the parameters, to minimize the gap between simulation and practical power system; this procedure greatly expands the availability of the dataset during faults with similar features as the field data;
3) The proposed data-driven method is validated using field data experiments; compared to widely adopted fundamental frequency phasor based based approaches, the proposed method only requires short data window of half a cycle during faults; compared to traveling wave based approaches, the proposed method is compatible with the typical sampling rate (several kHz) from Digital Fault Recorders (DFRs) in practical power systems.

The rest of this paper is organized as follows. Section II outlines the framework of the proposed method. Section III describes the generation process of the special dataset for training. Section IV shows the simulation and field data experiments. Section V conducts further discussions. Section VI concludes this paper.

## II. Framework of the Proposed Method

*A. Existing Data-Driven Methods and Limitations*

In the present literature, training and testing datasets are usually generated via simulation with typical system parameters, such as the typical equivalent source impedances, loading conditions, FIAs and fault resistances. Also, the field data are rarely utilized for training and testing. The limitations of traditional data-driven methods are explained as follows.

In fact, transmission line field data during faults in practical power systems are extremely "sparse". For transmission lines (voltage level ≥ 220kV) in State Grid Coorporation of China, the overall length of the line is about $6.2 \times 10^5$ km and the number of faults is about 2000 in year 2020 [21]. In other words, on average, there is only 1 fault per year for a 310 km transmission line.

With this extremely limited number of fault events in practical power systems, three possible training strategies are shown below. *First*, if the field data from only the interested line are utilized for training, the training sets are too small (only several fault events even during decades) to have reasonable fault location results. *Second*, if the field data from various lines are utilized for training, the training sets are still relatively small; in addition, the fault features of the interested line could be jeopardized: the mapping between the fault data and the fault location could be different for different lines. Moreover, if field data are utilized for training, the field data may not be able to cover the feature of any new fault event, with a specific FIA, fault location, fault type, fault resistance, equivalent source impedance, loading condition, and so on. *Third*, if simulation data are utilized for training, the mismatch between the simulation and practical power system will cause credibility issue of the simulation data. These strategies are summarized in Fig. 1. Therefore, with this small number of fault events in practical transmission lines, reasonable fault location results are hard to be obtained. The authors believe that this is the primary reason why the field data are rarely utilized for training and testing in the existing literature.

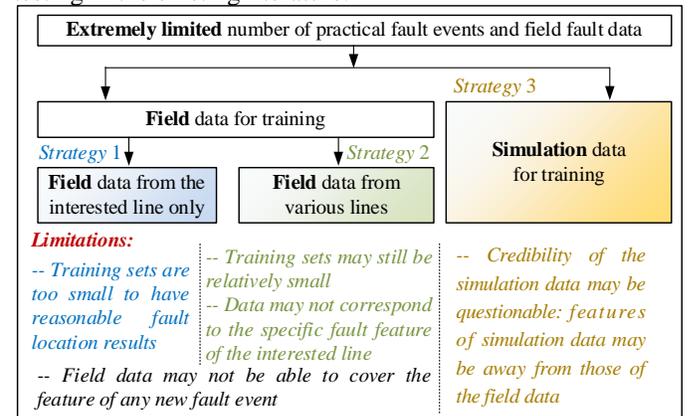

**Fig**.1 Possible strategies of traditional data-driven methods and the limitations

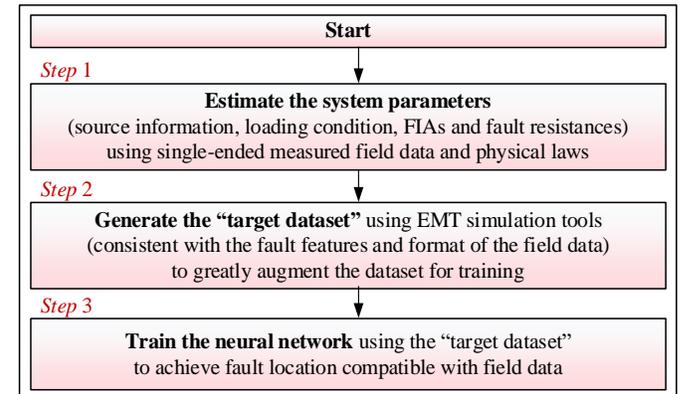

**Fig**.2 The Idea of the proposed physics-informed data-driven fault location

*B. Proposed Data-Driven Method*

Given the challenge of the application of traditional data-driven methods with field data, this paper proposes a physics-informed data-driven fault location approach. It only requires single-ended voltages and currents within half a cycle before and after the fault occurs. The proposed method includes three steps, as follows. *First*, the single-ended voltage and current measurements (field data) are used to estimate the system parameters, including source information, loading condition, FIAs and fault resistances. *Second*, the estimated parameters will be sent into the EMT simulation tools, to generate the "target dataset" (single-ended measurements) for training, which is consistent with the fault features and the format of the field data. *Third*, the target dataset is utilized to train the neural network, to achieve fault location using field data. In this case, the mismatch between the simulation and the field data is minimized. This idea is summarized in Fig. 2.

Next, the way of estimating the system parameters using single-ended measured field data is presented in section III. Details of the entire fault location procedure are shown in Section IV.

## III. ESTIMATION OF SYSTEM PARAMETERS USING LIMITED INFORMATION

Estimation of system parameters serves as the foundation to generate the target dataset. An example test system for data generation is shown in Fig. 3. The system consists of a transmission line, a fault within the line, and two equivalent sources at line terminals. With a given line of interest, the line parameters are typically available in the utility database. Therefore, for the EMT simulation of measurements at the local line terminal during the fault, the unknown system parameters include: (1) equivalent source impedances, (2) loading conditions, (3) FIAs and (4) fault resistances.

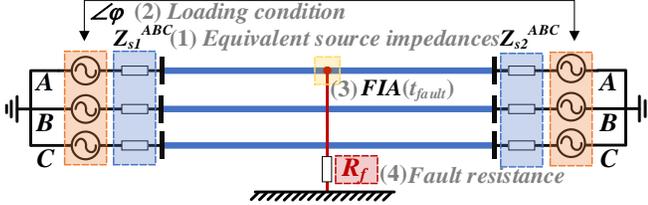

**Fig.3** Parameters to be estimated for transmission line system

Note that for the single-terminal fault location application, the above system parameters need to be estimated using single-ended measured field data within a limited time window before and during the fault (eg. the available time window during the fault could be only half a cycle, for transmission lines with fast-tripping techniques). With the limited information, the estimation of those parameters is quite challenging. In the rest of the section, the technical details of parameter estimation are presented below. Note that the parameter estimation method introduced in this section are with certain assumptions, which may not exactly match the real world scenarios. Consideration of those uncertainties will be covered during dataset generation and selection in section IV.

### A. Preparations

The parameter estimation method proposed in this paper is carried out in the $\alpha\beta0$ mode of the system. The relationship between the three mode components and the three phase components are shown in (1),

$$\begin{cases} \mathbf{u}^{\alpha\beta0} = \mathbf{T}_{Clarke}^{-1} \mathbf{u}^{ABC} \\ \mathbf{i}^{\alpha\beta0} = \mathbf{T}_{Clarke}^{-1} \mathbf{i}^{ABC} \end{cases}, \begin{cases} \mathbf{R}^{\alpha\beta0} = \mathbf{T}_{Clarke}^{-1} \mathbf{R}^{ABC} \mathbf{T}_{Clarke} \\ \mathbf{L}^{\alpha\beta0} = \mathbf{T}_{Clarke}^{-1} \mathbf{L}^{ABC} \mathbf{T}_{Clarke} \end{cases} \quad (1)$$

where $\mathbf{T}_{Clarke}^{-1} = [2/3, -1/3, -1/3; 0, 1/\sqrt{3}, -1/\sqrt{3}; 1/3, 1/3, 1/3]$. The variables $(\cdot)^{\alpha\beta0}$ and $(\cdot)^{ABC}$ mean the three mode component and the three phase component of $(\cdot)$ in the matrix form, where $(\cdot)$ represents the variable $\mathbf{u}, \mathbf{i}, \mathbf{R},$ and $\mathbf{L}$.

The fault initiation time $t_f$ is captured by detection of the sudden change of the derivative of the terminal mode current $i_1^j(t)$, where $j = \alpha, \beta,$ and $0$ represent the three modes, and the subscript 1 represents the variable at the local terminal. The derivative of the current $i_1^j$ is numerically calculated with the central difference method $i_1^{j(1)} = (i_1^j(t+\Delta t) - i_1^j(t-\Delta t))/2\Delta t$, where $\Delta t$ is the sampling interval. Afterwards, the fault initiation time $t_f$ is determined as the first time index satisfying the following equation,

$$t_f = t \big|_{i_1^{j(1)} > k_{tf} I_1^{j(1)}{}_{max}} \quad (2)$$

where $I_1^{j(1)}{}_{max}$ is the maximum value of the derivative of $i_1^{j(1)}$ during the pre-fault steady state, and $k_{tf}$ is a user-defined coefficient larger than 1 (eg. $k_{tf}$ = 1.5). The following calculations are based on the time reference of $t_f$, i.e., the system time is shifted to ensure $t_f = 0$ sec.

### B. Equivalent source impedances

The equivalent model of the local terminal three phase equivalent source is shown in Fig. 4. Here the model is built in the mode domain. $u_{s1}^j$ is the equivalent terminal voltage, $i_{j1}$ is the terminal current, $R_{s1}^j$, and $L_{s1}^j$ are the equivalent source resistance and inductance, where $j = \alpha, \beta,$ and $0$ (same definition in the rest of the paper).

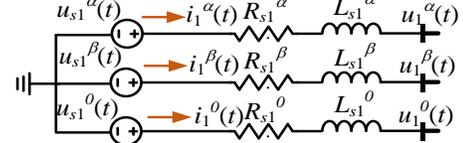

**Fig.4** Equivalent source model for three phase transmission lines

For simplicity, the equivalent source impedances for two terminal sources are assumed to be identical and are estimated with the single-ended measurements in each mode. From Fig. 4, the physics law of the circuit at any given time $t$ can be written as (3a). Substitute $t$ as $t - \Delta T$ in (3a), the physics law of the circuit at $t - \Delta T$ can be written as (3b), where $\Delta T$ is time of one cycle of the fundamental frequency,

$$u_{s1}^j(t) - u_1^j(t) = R_{s1}^j i_1^j(t) + L_{s1}^j di_1^j(t)/dt \quad (3a)$$

$$u_{s1}^j(t-\Delta T) - u_1^j(t-\Delta T) = R_{s1}^j i_1^j(t-\Delta T) + L_{s1}^j di_1^j(t-\Delta T)/dt \quad (3b)$$

In (3), The terminal voltage $u_1^j(t)$ and current $i_1^j(t)$ are available measurements. However, the source voltage $u_{s1}^j(t)$ is unknown. To eliminate the source voltages, the difference between (3a) and (3b) corresponds to the fault component,

$$u_1^j(t) - u_1^j(t-\Delta T) = R_{s1}^j \left[ i_1^j(t) - i_1^j(t-\Delta T) \right] \\ + L_{s1}^j d \left[ i_1^j(t) - i_1^j(t-\Delta T) \right]/dt \quad (4)$$

Considering the practical scenarios, here we assume that the available time window during faults is at least half a cycle. In this case, the range of $t$ is $t \in [t_f, t_f + \Delta T/2]$ for (4), where $t_f$ is the fault initiation time solved in Section III.A. To solve the two unknowns $R_{s1}^j$ and $L_{s1}^j$, the derivative of the terminal current in (4) is discretized with the central difference method. Therefore, equation (4) becomes a linear equation,

$$\left[ \Delta \mathbf{i}_1^j \quad \Delta \mathbf{i}_1^{j(1)} \right] \mathbf{x} = \left[ \Delta \mathbf{u}_1^j \right] \quad (5)$$

where $\Delta \mathbf{y} = [y(t_f) - y(t_f - \Delta T), \ y(t_f + \Delta t) - y(t_f + \Delta t - \Delta T), \ y(t_f + 2\Delta t) - y(t_f + 2\Delta t - \Delta T), \ ... \ , \ y(t_f + \Delta T/2) - y(t_f - \Delta T/2)]^T$ is a vector, $\mathbf{y}$ represents $\mathbf{u}_1^j, \mathbf{i}_1^j, \mathbf{i}_1^{j(1)}$, and $\mathbf{x} = [R_{s1}^j, L_{s1}^j]^T$.

Therefore, the mode $j$ source resistance and inductance can be solved with the least square scheme,

$$\mathbf{x} = ([\Delta \mathbf{i}_1^j \quad \Delta \mathbf{i}_1^{j(1)}]^T [\Delta \mathbf{i}_1^j \quad \Delta \mathbf{i}_1^{j(1)}])^{-1} [\Delta \mathbf{i}_1^j \quad \Delta \mathbf{i}_1^{j(1)}]^T [\Delta \mathbf{u}_1^j] \quad (6)$$

### C. Loading conditions

In this section, the estimation of the loading condition is achieved with the phasor domain model before the fault. The last zero crossing point of the terminal voltage measurement $u_1^j(t)$ before the fault initiation time $t_f$ is defined as $t_0$, where $u_1^j(t)$ is increasing around $t_0$. Afterwards, the local-end voltage phasor $\tilde{U}_1^j$ and current phasors $\tilde{I}_1^j$ are calculated with the time domain measurements within the data window of $[t_0-T, t_0]$, where $T$ is one period (e.g. 0.02 s for 50 Hz system).

With the modeling in phasor domain, the loading condition is defined as the difference of voltage source phase angles between the remote and the local side. This phase angle difference is calculated as follows. With the calculated local-end equivalent source impedance in Section III.B. and the assumption of identical equivalent source impedances for two terminal sources, equation (7) can be listed according to the KVL between two sources in the steady phasor domain,

$$\begin{cases} \tilde{U}_{s1}{}^{j} = \tilde{U}_1{}^{j} + \tilde{I}_1{}^{j} Z_{s1}{}^{j} \\ \tilde{U}_{s2}{}^{j} = \tilde{U}_1{}^{j} - \tilde{I}_1{}^{j}(Z_{s1}{}^{j} + Z_l{}^{j}) \end{cases} \quad (7)$$

where $\tilde{U}_{sk}{}^{j}$ is the phasor of equivalent source voltage represents the local terminal ($k = 1$) and the remote terminal ($k = 2$) respectively (same definition of the subscript $k$ for the rest of the paper). $Z_{s1}{}^{j} = R_{s1}{}^{j} + j\omega L_{s1}{}^{j}$ and $Z_l{}^{j}$ are the equivalent source impedance and line impedance for entire length of the line, where $R_{s1}{}^{j}$ and $L_{s1}{}^{j}$ are solved in Section III.B, and $Z_l{}^{j}$ is the known line parameter.

### D. Fault inception angles

In this section, the method of calculating the fault inception angle (FIA) is proposed. First, the fault initiation time $t_f$ is calculated in (2). With the zero-crossing time $t_0$ in section III.C, the fault inception angle is determined as,

$$FIA = 2\pi(t_f - t_0)/\Delta T \quad (8)$$

Note that the value $t_f$ is actually the time when the current measurements shows sudden changes due to the fault occurs. Therefore, the actual time when the fault occurs is earlier than $t_f$, where the traveling wave propagation time from the fault location to the local terminal should also be considerd. Here the time of traveling wave propagation is neglected, and the estimation error of the fault initiation time is small when the line is not quite long (12 degrees error for 200km lines).

### E. Fault resistances

In this section, the possible range of the fault resistance is estimated. Since the available time window during the fault is only half a cycle, phasor domain methods may not be suitable; in this case, the estimation can be achieved by solving the time domain faulted line model.

#### 1) Formulation of the fault resistance estimation

For simplicity, the transmission lines at both sides of the fault are modeled as R-L models. The faulted transmission line model after Clarke transformation is described as,

$$\boldsymbol{u}_{s1}{}^{\alpha\beta0}(t) - \boldsymbol{u}_f{}^{\alpha\beta0}(t) = \boldsymbol{R}_{eq1}{}^{\alpha\beta0}\boldsymbol{i}_1{}^{\alpha\beta0}(t) + \boldsymbol{L}_{eq1}{}^{\alpha\beta0}d\boldsymbol{i}_1{}^{\alpha\beta0}(t)/dt$$

$$\boldsymbol{u}_{s2}{}^{\alpha\beta0}(t) - \boldsymbol{u}_f{}^{\alpha\beta0}(t) = \boldsymbol{R}_{eq2}{}^{\alpha\beta0}\boldsymbol{i}_2{}^{\alpha\beta0}(t) + \boldsymbol{L}_{eq2}{}^{\alpha\beta0}d\boldsymbol{i}_2{}^{\alpha\beta0}(t)/dt \quad (9)$$

$$\boldsymbol{i}_f{}^{\alpha\beta0}(t) = \boldsymbol{i}_1{}^{\alpha\beta0}(t) + \boldsymbol{i}_2{}^{\alpha\beta0}(t) \quad \boldsymbol{i}_f{}^{\alpha\beta0}(t) = \boldsymbol{Y}_f{}^{\alpha\beta0}\boldsymbol{u}_f{}^{\alpha\beta0}(t)$$

where subscript '$f$' represents the variables within the faulted branch (the same definition is used in the rest of the paper). $(\cdot)^{\alpha\beta0}$ is calculated with $(\cdot)^{ABC}$ after Clarke transformation. For $k = 1$ or 2, $\boldsymbol{R}_{eqk}{}^{ABC} = \boldsymbol{R}_{sk}{}^{ABC} + \boldsymbol{R}_{lk}{}^{ABC}$, where $\boldsymbol{R}_{sk}{}^{ABC}$ is the equvalent source resistance, $\boldsymbol{R}_{lk}{}^{ABC}$ is the equvalent line resistance at the left side ($k = 1$) and the right side ($k = 2$) of the fault. With the fault location $l_f$, $\boldsymbol{R}_{l1}{}^{ABC} = l_f/l_{line}\boldsymbol{R}_l{}^{ABC}$, $\boldsymbol{R}_{l2}{}^{ABC} = (1 - l_f/l_{line})\boldsymbol{R}_l{}^{ABC}$, and $\boldsymbol{R}_l{}^{ABC}$ is the resistance matrix of the entire line. Inductance variables are similarly defined by replacing $\boldsymbol{R}$ with $\boldsymbol{L}$.

Table. I Value of the fault matrix $\boldsymbol{Y}_f{}^{\alpha\beta0}$

| SLG (AG) | LL (BC) | LLG (BCG) | 3PH (ABC) |
|---|---|---|---|
| $\begin{bmatrix} \frac{2}{3}Y_f & 0 & \frac{2}{3}Y_f \\ 0 & 0 & 0 \\ \frac{1}{3}Y_f & 0 & \frac{1}{3}Y_f \end{bmatrix}$ | $\begin{bmatrix} 0 & 0 & 0 \\ 0 & 2Y_f & 0 \\ 0 & 0 & 0 \end{bmatrix}$ | $\begin{bmatrix} \frac{1}{3}Y_f & 0 & -\frac{2}{3}Y_f \\ 0 & Y_f & 0 \\ -\frac{1}{3}Y_f & 0 & \frac{2}{3}Y_f \end{bmatrix}$ | $\begin{bmatrix} 3Y_f & 0 & 0 \\ 0 & 3Y_f & 0 \\ 0 & 0 & 0 \end{bmatrix}$ |

For different fault types, the value of the fault matrix $\boldsymbol{Y}_f{}^{\alpha\beta0}$ in (9) are shown in Table I, where $Y_f$ is the fault conductance, and $R_f = 1/Y_f$ is the fault resistance.

From Table I, for LL (line to line), LLG (double line to ground), and 3PH (3 phase) faults, only $\beta$ mode is needed for network construction, with details shown in section III.E.2). For SLG (single line to ground) faults, both $\alpha$ mode and 0 mode are used for network construction, with details shown in section III.E.3). Note that for each fault group (SLG, LL, LLG), one specific fault type is selected as an example, and other fault types can similarly be derived by rotating the phase sequence.

#### 2) Faulted line model during LL, LLG, 3PH faults

Take the LL fault (BC fault) as an example. Equation (9) can be simplified into (10). One can replace $2Y_f$ with $Y_f$ and $3Y_f$ in (10) to obtain the model during LLG and 3PH faults, respectively.

$$u_{s1}{}^{\beta}(t) - u_f{}^{\beta}(t) = R_{eq1}{}^{\beta}i_1{}^{\beta}(t) + L_{eq1}{}^{\beta}di_1{}^{\beta}(t)/dt$$

$$u_{s2}{}^{\beta}(t) - u_f{}^{\beta}(t) = R_{eq2}{}^{\beta}i_2{}^{\beta}(t) + L_{eq2}{}^{\beta}di_2{}^{\beta}(t)/dt \quad (10)$$

$$i_1{}^{\beta}(t) + i_2{}^{\beta}(t) = i_f{}^{\beta}(t) \quad i_f{}^{\beta}(t) = 2Y_f u_f{}^{\beta}(t)$$

where $R_{eq\beta1}$, $L_{eq\beta1}$, $R_{eq\beta2}$, $L_{eq\beta2}$ are the second element of the diagonal vector in matrices $\boldsymbol{R}_{eq1}{}^{\alpha\beta0}$, $\boldsymbol{L}_{eq1}{}^{\alpha\beta0}$, $\boldsymbol{R}_{eq2}{}^{\alpha\beta0}$, $\boldsymbol{L}_{eq2}{}^{\alpha\beta0}$, respectively. $u_{s1}{}^{\beta}$, $u_{s2}{}^{\beta}$, $i_1{}^{\beta}$, $i_2{}^{\beta}$, $u_f{}^{\beta}$, $i_f{}^{\beta}$ are the second element of the vector $\boldsymbol{u}_{s1}{}^{\alpha\beta0}$, $\boldsymbol{u}_{s2}{}^{\alpha\beta0}$, $\boldsymbol{i}_1{}^{\alpha\beta0}$, $\boldsymbol{i}_2{}^{\alpha\beta0}$, $\boldsymbol{u}_f{}^{\alpha\beta0}$, $\boldsymbol{i}_f{}^{\alpha\beta0}$. Equation (10) is equivalent to a circuit as shown in Fig. 5, where $j = \beta$, $R_2 = R_{eq1}{}^{\beta}$, $L_2 = L_{eq1}{}^{\beta}$, $R_3 = R_{eq2}{}^{\beta}$, $L_3 = L_{eq2}{}^{\beta}$. $R_4 = 1/(2Y_f)$, and $L_4 = 0$.

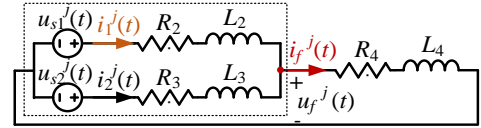

**Fig**. 5. Single mode equivalent circuit to solve the terminal current

From Fig. 5 and equation (10), with $L_4 = 0$, the constructed faulted line model is a second order model in general. To further simplify the analyzing procedure, the circuit in the dashed box can be simplified as a Thevenin's branch according to the Appendix. To solve $i_f{}^{j}(t)$, the overall circuit can be simplified as Fig. 6, where $j = \beta$, $R_1 = R_{eq}{}^{\beta} + 1/(2Y_f)$, $L_1 = L_{eq}{}^{\beta}$. $u_{seq}{}^{\beta}(t) = (R_{eq2}{}^{\beta}u_{s1}{}^{\beta}(t) + R_{eq1}{}^{\beta}u_{s2}{}^{\beta}(t))/(R_{eq1}{}^{\beta} + R_{eq2}{}^{\beta})$, $R_{eq}{}^{\beta} = (R_{eq1}{}^{\beta}R_{eq2}{}^{\beta})/(R_{eq1}{}^{\beta} + R_{eq2}{}^{\beta})$, and $L_{eq}{}^{\beta} = (L_{eq1}{}^{\beta}L_{eq2}{}^{\beta})/(L_{eq1}{}^{\beta} + L_{eq2}{}^{\beta})$.

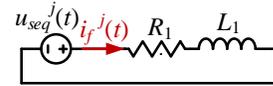

**Fig**. 6. Single mode equivalent circuit to solve the fault current

From Fig. 6, $i_f{}^{j}(t)$ satisfies a first order ordinary differential equation (ODE). $i_f{}^{j}(t)$ can be obtained by solving,

$$di_f{}^{j}(t)/dt + B_1 i_f{}^{j}(t) = B_2(t) \quad (14)$$

where $B_1 = R_1/L_1$, $B_2(t) = u_{seq}{}^{j}(t)/L_1$, and $j = \beta$. The initial condition is $i_f{}^{j}(0_+) = 0$.

Next, from the result of $i_f{}^{j}(t)$ and Fig. 5, $i_1{}^{j}(t)$ also satisfies a first order ODE,

$$di_1{}^{j}(t)/dt + B_3 i_1{}^{j}(t) = B_4(t) \quad (15)$$

where $B_3 = R_2/L_2$, $B_4(t) = [u_{s1}{}^{j}(t) - R_4 i_f{}^{j}(t) - L_4 di_f{}^{j}(t)/dt]/L_2$, and $j = \beta$. The initial condition $i_1{}^{j}(0_+)$ is known from measurements.

To sum up, $i_1{}^{j}(t)$ can be obtained by solving (14) and (15).

#### 3) Faulted line model during SLG fault

This section further derives the line model during SLG (e.g AG fault). Equation (9) can be simplified into (16).

$$u_{s1}^{\alpha}(t) - u_{f}^{\alpha}(t) = R_{eq1}^{\alpha} i_{1}^{\alpha}(t) + L_{eq1}^{\alpha} di_{1}^{\alpha}(t)/dt$$
$$u_{s2}^{\alpha}(t) - u_{f}^{\alpha}(t) = R_{eq2}^{\alpha} i_{2}^{\alpha}(t) + L_{eq2}^{\alpha} di_{2}^{\alpha}(t)/dt$$
$$-u_{f}^{0}(t) = R_{eq1}^{0} i_{1}^{0}(t) + L_{eq1}^{0} di_{1}^{0}(t)/dt$$
$$-u_{f}^{0}(t) = R_{eq2}^{0} i_{2}^{0}(t) + L_{eq2}^{0} di_{2}^{0}(t)/dt \quad (16)$$
$$i_{f}^{\alpha}(t) = 2Y_{f}/3 \cdot u_{f}^{\alpha}(t) + 2Y_{f}/3 \cdot u_{f}^{0}(t)$$
$$i_{f}^{0}(t) = Y_{f}/3 \cdot u_{f}^{\alpha}(t) + Y_{f}/3 \cdot u_{f}^{0}(t)$$
$$i_{f}^{\alpha}(t) = i_{1}^{\alpha}(t) + i_{2}^{\alpha}(t) \quad i_{f}^{0}(t) = i_{1}^{0}(t) + i_{2}^{0}(t)$$

The 3rd and 4th row of (16) represent two Thevenin's branches in parallel. From the Appendix, the two branches can be simplified as a Thevenin's branch ($R_{eq}^{0}$ and $L_{eq}^{0}$). After eliminating the 0 mode variables in (16),

$$u_{s1}^{\alpha}(t) - u_{f}^{\alpha}(t) = R_{eq1}^{\alpha} i_{1}^{\alpha}(t) + L_{eq1}^{\alpha} di_{1}^{\alpha}(t)/dt$$
$$u_{s2}^{\alpha}(t) - u_{f}^{\alpha}(t) = R_{eq2}^{\alpha} i_{2}^{\alpha}(t) + L_{eq2}^{\alpha} di_{2}^{\alpha}(t)/dt \quad (17)$$
$$u_{f}^{\alpha}(t) = 1/2 \cdot R_{eq}^{0} i_{f}^{\alpha}(t) + 1/2 \cdot L_{eq}^{0} di_{f\alpha}(t)/dt + 3/2Y_{f} \cdot i_{f}^{\alpha}(t)$$
$$i_{f}^{\alpha}(t) = i_{1}^{\alpha}(t) + i_{2}^{\alpha}(t)$$

where $R_{eq}^{0} = R_{eq1}^{0} R_{eq2}^{0}/(R_{eq1}^{0} + R_{eq2}^{0})$, $L_{eq}^{0} = L_{eq1}^{0} L_{eq2}^{0}/(L_{eq1}^{0} + L_{eq2}^{0})$.

Equation (17) is also equivalent to the circuit in Fig. 5, where $j = \alpha$, $R_2 = R_{eq1}^{\alpha}$, $L_2 = L_{eq1}^{\alpha}$, $R_3 = R_{eq2}^{\alpha}$, $L_3 = L_{eq2}^{\alpha}$, $R_4 = 1/2R_{eq}^{0} + 3/2Y_f$, $L_4 = 1/2L_{eq}^{0}$. Similarly, Fig. 5 can be further simplified as Fig. 6, where $j = \alpha$, $R_1 = R_{eq}^{\alpha} + 1/2R_{eq}^{0} + 3/2Y_f$, $L_1 = L_{eq}^{\alpha} + 1/2L_{eq}^{0}$, $u_{seq}^{\alpha}(t) = (R_{eq2}^{\alpha} u_{s1}^{\alpha}(t) + R_{eq1}^{\alpha} u_{s2}^{\alpha}(t))/(R_{eq1}^{\alpha} + R_{eq2}^{\alpha})$, $R_{eq}^{\alpha} = (R_{eq1}^{\alpha} R_{eq2}^{\alpha})/(R_{eq1}^{\alpha} + R_{eq2}^{\alpha})$, and $L_{eq}^{\alpha} = (L_{eq1}^{\alpha} L_{eq2}^{\alpha})/(L_{eq1}^{\alpha} + L_{eq2}^{\alpha})$. Similarly, $i_1^j(t)$ can be obtained by solving (14) and (15).

### 4) General solution of the faulted line model

According to Section.E.2 and Section.E.3, different types of fault networks can be represented by (14) and (15). Next, the analytical solution of (14) is first derived to obtain the fault current $i_f^j(t)$. For (14), $B_1 = R_1/L_1$ is a constant, and $B_2(t)$ is,

$$B_2(t) = M_{step1} \cos(\omega t + \varphi_{step1}) \quad (18)$$

where $M_{step1} = \sqrt{2}|\tilde{U}_{seq}^j|/L_1$, $\varphi_{step1}$ is the phase angle of $\tilde{U}_{seq}^j$, $\omega$ is the rated angular frequency, and $\tilde{U}_{seq}^j$ is the phasor representation of $u_{seq}^j(t)$. The solution of (14) in time domain is,

$$i_f^j(t) = -M_{sol1} \cos(\varphi_{sol1}) e^{-B_1 t} + M_{sol1} \cos(\omega t + \varphi_{sol1}) \quad (19)$$

where $M_{sol1} = \sqrt{a_1^2 + b_1^2 + 2a_1 b_1 \cos(\varphi_{a1} - \varphi_{b1})}$, $\varphi_{a1} = \varphi_{step1}$, $\varphi_{b1} = \varphi_{step1} - \frac{\pi}{2}$, $\varphi_{sol1} = \arctan \frac{a_1 \sin \varphi_{a1} + b_1 \sin \varphi_{b1}}{a_1 \cos \varphi_{a1} + b_1 \cos \varphi_{b1}}$, $a_1 = \frac{B_1 M_{step1}}{B_1^2 + \omega^2}$, $b_1 = \frac{\omega M_{step1}}{B_1^2 + \omega^2}$.

With the result of $i_f^j(t)$ in (19), equation (15) can be formulated, where $B_3 = R_2/L_2$ is a constant, and $B_4(t)$ is,

$$B_4(t) = M_{step2} \cos(\omega t + \varphi_{step2}) + A_{step2} e^{-t/\tau_{step2}} \quad (20)$$

where $M_{step2} = \sqrt{2}|\tilde{U}_{s1}^j - (R_4 + j\omega L_4)\tilde{I}_f^j|/L_2$, $\varphi_{step2}$ is the phase angle of $\tilde{U}_{s1}^j - (R_4 + j\omega L_4)\tilde{I}_f^j$, and $\tilde{I}_f^j = M_{sol1} \angle \varphi_{sol1}/\sqrt{2}$ is the sinusoidal part of $i_f^j(t)$ in (19). $A_{step2} = (R_4 M_{sol1} \cos \varphi_{sol1} - B_1 L_4 M_{sol1} \cos \varphi_{sol1})/L_2$.

The solution of (20) in time domain is,

$$i_1^j(t) = [i_1^j(0_+) - M_{sol2} \cos(\varphi_{sol2}) - A_{step2}/(B_3 - B_1)] e^{-B_3 t}$$
$$+ M_{sol2} \cos(\omega t + \varphi_{sol2}) + A_{step2}/(B_3 - B_1) e^{-B_1 t} \quad (21)$$

where $M_{sol2} = \sqrt{a_2^2 + b_2^2 + 2a_2 b_2 \cos(\varphi_{a2} - \varphi_{b2})}$, $\varphi_{a2} = \varphi_{step2}$, $\varphi_{b2} = \varphi_{step2} - \frac{\pi}{2}$, $\varphi_{sol2} = \arctan \frac{a_2 \sin \varphi_{a2} + b_2 \sin \varphi_{b2}}{a_2 \cos \varphi_{a2} + b_2 \cos \varphi_{b2}}$, $a_2 = \frac{B_3 M_{step2}}{B_3^2 + \omega^2}$, $b_2 = \frac{\omega M_{step2}}{B_3^2 + \omega^2}$.

### 5) Range of fault resistance

To sum up, the current at the local terminal $i_1^j(t)$ can be represented with various parameters including the fault resistance $R_f = 1/Y_f$ and the fault location $l_f$, with details shown in Table II. Therefore, with the actual measured $i_1^j(t)$ within $t_f \le t \le t_f + \Delta T/2$, the possible range of $R_f$ can be determined. For simplification, here the maximum value of $|i_1^j(t)|$ is taken as the criterion. Since $l_f$ is unknown, the range of $R_f$ is selected such that the theoretical and measured maximum values of $|i_1^j(t)|$ are consistent with varying $l_f$, i.e.,

$$(1-c) \cdot \max|i_{1,meas}^j(t)| \le \max_{R_f} |i_1^j(t)| \le (1+c) \cdot \max|i_{1,meas}^j(t)| \quad (22)$$

where $\max|i_{1,meas}^j(t)|$ is a constant from measurements, $t_f \le t \le t_f + \Delta T/2$, $0 \le l_f \le l_{line}$, and $c$ is a user-defined margin to consider parameter and measurement uncertainties (typical value: 5%). Here $j = \alpha$ for SLG faults and $j = \beta$ for other types of faults.

Table II Solution Coefficient

| Type | SLG | LL | LLG | 3PH |
|---|---|---|---|---|
| $u_{seq}^j$ | $u_{seq}^\alpha$ | $u_{seq}^\beta$ | $u_{seq}^\beta$ | $u_{seq}^\beta$ |
| $u_{s1}^j$ | $u_{s1}^\alpha$ | $u_{s1}^\beta$ | $u_{s1}^\beta$ | $u_{s1}^\beta$ |
| $R_1$ | $R_{eq}^\alpha + 1/2R_{eq}^0 + 3/2Y_f$ | $R_{eq}^\beta + 1/2Y_f$ | $R_{eq}^\beta + 1/Y_f$ | $R_{eq}^\beta + 1/3Y_f$ |
| $L_1$ | $L_{eq}^\alpha + 1/2L_{eq}^0$ | $L_{eq}^\beta$ | $L_{eq}^\beta$ | $L_{eq}^\beta$ |
| $R_2$ | $R_{eq1}^\alpha$ | $R_{eq1}^\beta$ | $R_{eq1}^\beta$ | $R_{eq1}^\beta$ |
| $L_2$ | $L_{eq1}^\alpha$ | $L_{eq1}^\beta$ | $L_{eq1}^\beta$ | $L_{eq1}^\beta$ |
| $R_4$ | $1/2R_{eq}^0 + 3/2Y_f$ | $1/2Y_f$ | $1/Y_f$ | $1/3Y_f$ |
| $L_4$ | $1/2L_{eq}^0$ | 0 | 0 | 0 |

## IV. PHYSICS-INFORMED DATA-DRIVEN FAULT LOCATION

After the fault occurs, the fault location will be estimated using single-ended measurements. This section will introduce the procedure of the proposed physics-informed data-driven fault location method. The flow chart of the proposed method is depicted in Fig. 7.

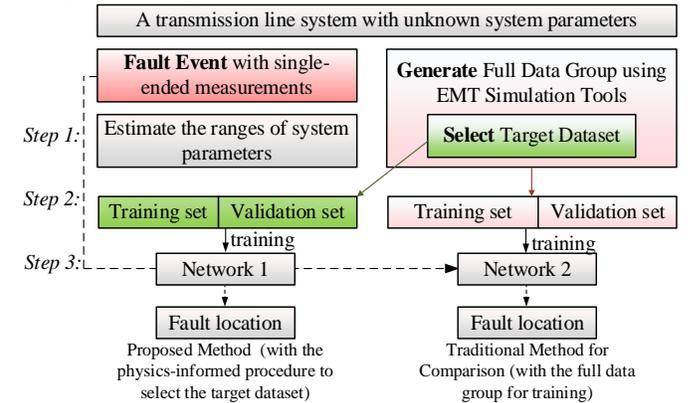

**Fig. 7.** Flow chart of the proposed method and method for comparison

*Step 1*: With the single-ended measurements from the fault event, the ranges of system parameters are estimated according to section III.

*Step 2*: The target dataset is generated via EMT simulation based on the ranges of system parameters. The features of the fault event are well described by the target dataset. To reduce the computational complexity of the proposed method, the target dataset is not generated by running extensive EMT simulations every time a fault event is fed into the fault location method. Instead, according to the ranges of system parameters, the target dataset is selected from the "full data group". The full data group covers a large number of possible combinations of system parameters (equivalent source impedances, loading conditions, FIAs, fault resistances and fault location), and is generated offline before the fault location procedure. Since the target dataset also covers a certain range of system parameters, the uncertainties of th estimated system parameters will not affect much the performance of the data driven approach. In this paper, Matlab Simulink is utilized for EMT simulation.

*Step 3*: The target dataset is then fed into the data driven network for training. There are various possible candidates for the data driven network, including NN, GNN, DTR, CNN, among others. In this paper, for simplification, the classic NN is adopted to verify the idea of the proposed physics-informed data-driven fault location method.

To verify the importance of selecting the target dataset for training, the method for comparison is also introduced in Fig. 7. It directly utilizes the full data group for training, without the physics-informed procedure to select the target dataset.

## V. SIMULATION RESULTS

In this section, various fault events in an example transmission line system are studied to validate the proposed fault location method. First, with the example transmission line system, the way to generate the full data group is presented. Second, various fault events for validation are introduced. Finally, the performances of the proposed method are evaluated during example fault events.

### A. Example Test System and Full Data Group Generation

The dual-terminal transmission line simulation system is shown in Fig. 8. The available measurements are the local terminal instantaneous voltages and currents with a sampling rate of 80 samples per cycle (4k samples per second for 50 Hz system) according to IEC 61850-SV standard. The required parameters to construct the simulation system include fault type, line parameters, equivalent source impedances, loading conditions, FIAs, fault resistances and fault location. The system voltage level is 500kV. The line is with the length of 200 km; other line parameters are shown in Table. III. and the entire line length is 200km.

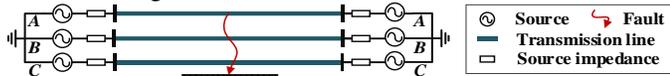

Fig. 8. Three phase transmission line simulation model

Table. III Simulation line parameters

| $R$(ohm/km) | $R_s = 0.106$, $R_m = 0.091$ | Subscripts "s" and "r" represent self and mutual parameters |
|---|---|---|
| $L$(H/km) | $L_s = 0.0016$, $L_m = 0.0008$ | |
| $C$(F/km) | $C_s = 0.129 \times 10^{-7}$, $C_m = -0.025 \times 10^{-7}$ | |

To generate the full data group, the rest of the system parameters will be assigned as typical values, which are shown in Table. IV. The full data group is generated for each fault type (SLG, LL, LLG and 3PH faults). Faults with high fault resistances will only be included during SLG faults. For each system parameter, the number candidates are limited, to minimize the size of the full data group. Even so, for A-G fault as an example, the present full data group still consists of over 1 million (1,260,864) fault events. For each fault event, the measurements of local terminal instantaneous currents ($i_1^A(t)$, $i_1^B(t)$, $i_1^C(t)$) and voltages ($u_1^A(t)$, $u_1^B(t)$, $u_1^C(t)$) half cycle before and after the fault occurs are resized into a matrix: [$i_1^A(t)$, $i_1^B(t)$, $i_1^C(t)$, $u_1^A(t)$, $u_1^B(t)$, $u_1^C(t)$]. Therefore, each set of data is with the dimension of 81x6.

Table. IV Parameters in the full data group, simulation

| Parameters | Value |
|---|---|
| Source impedance/ohm | $Z_{\alpha}=Z_{\beta}=$1+j5, 2+j8, 5+j10<br>$Z_0=$3+j15, 5+j10, 7+j15 |
| Loading condition/deg | ± 2, ± 6, ± 10, ± 12 |
| Fault resistance/ohm | 0.5, 2.5, 4.5, 6.5, 8.5, 15, 35, (low resistance)<br>50, 100, 200, 300 (high resistance) |
| Fault location/km | 1, 2, 3, 4, …, 198, 199 (step: 1 km) |
| FIA/deg | 0, 45, 90, …, 270, 315 (step: 45 deg) |

### B. Various Fault Events to Formulate the Testing Set

To ensure the reliability of the proposed method, the system parameters in the testing set (to generate various fault events) will be different from those in the full data group. Table. V shows the parameters in the testing set for each fault type (SLG, LL, LLG and 3PH faults), where the high fault resistance will only be considered during SLG faults.

Table. V Parameters in the testing set, simulation

| Parameters | Value |
|---|---|
| Source impedance/ohm | $Z_{s\alpha 1}=Z_{s\beta 1}=$2.4+j6.6  $Z_{s\alpha 2}=Z_{s\beta 2}=$2.9+j7<br>$Z_{s01}=$4.1+j12.3 $Z_{s02}=$5.2+j16.4 |
| Loading condition/deg | 12 |
| Fault resistance/ohm | 0.1, 1, 3, 7, (low resistance)<br>40, 80, 160, 320 (high resistance) |
| Fault location/km | 25, 50, 75, 100, 125, 150, 175 |
| FIA/deg | 67.5 |

### C. Fault Location for an Example Low Resistance Fault

Take a 1ohm A-G fault at 25 km as an example. The local terminal measurements are shown in Fig. 9.

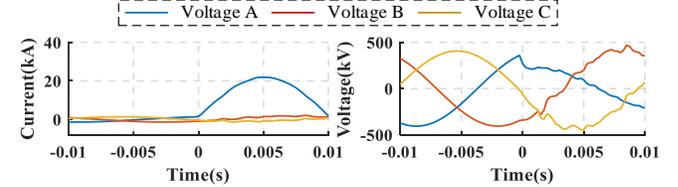

Fig. 9. Current and voltage measurements, 1 ohm A-G fault at 25 km

*Step 1*: According to section III, the system parameters are estimated below. From (6), the *source impedance* is calculated as $Z_{s\alpha 1}$ = 2.3977+j6.6901, $Z_{s01}$ = 4.1004+j12.3902 ohm. From (7), the *loading condition* is calculated as 11.9211 deg. From (8), the *fault inception angle* is calculated as 67.5 deg. From (22), the calculation procedure of the *range of fault resistance* is illustrated in Fig. 10. From the current measurements, the value of $\max|i_{1,meas}^\alpha(t)|$ is 14.63 kA. With the typical value of $c = 5\%$, upper bound is $(1+c)*\max|i_{1,meas}^\alpha(t)|$ = 15.36kA and the lower bound is $(1-c)*\max|i_{1,meas}^\alpha(t)|$ = 13.90 kA. With varying fault resistance and fault location, the value of $\max|i_1^\alpha(t)|$ is also depicted as the "maximum current surface" in Fig. 10. From the figure, the range of fault resistance is $R_f \in [0, 7.7] ohm$.

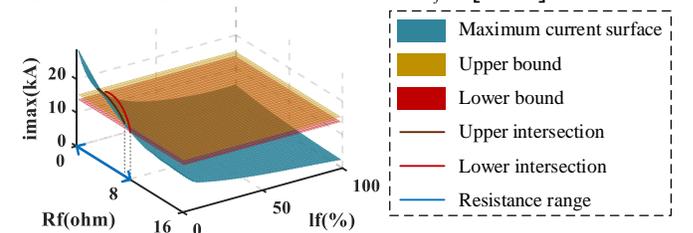

Fig. 10 The range of the fault resistance, 1 ohm A-G fault at 25 km

*Step 2*: Within the full data group, the target dataset is selected according to the estimated system parameters in *Step 1*. The target dataset is the minimum subset of the full data group that could approximately cover the range of the estimated system values. The details are shown in Table. VI.

Step 3: The target dataset is adopted to train the NN. As mentioned before, the dimension of the input data is 81×6. Reshape the input data into a vector with the dimension of

486×1. The network has 5 hidden layers with the size of 256, 128, 64, 32, and 16, respectively. The output layer is with the dimension of 1×1. The loss function is L2 loss. The learning rate is 0.001, the batch size is 128, and the epoch is 70. The target dataset is randomly devided into 80% for training and 20% for validation.

**Table**.VI Parameters in the target dataset, simulation

| Parameters | Value |
|---|---|
| Source impedance/ohm | $Z_{s\alpha 1} = 1+j5, 2+j8$; $Z_{s01} = 3+j15, 5+j10$ |
| Loading condition/deg | 10, 12 |
| Fault resistance/ohm | 0.5, 2.5, 4.5, 6.5, 8.5 |
| Fault location/km | 1, 2, 3, 4, …, 198, 199 (step: 1 km) |
| FIA/deg | 45, 90 |

During the network training, to avoid the randomness for each training (selection of training/validation datasets, batches, initial values, etc.), this paper adopts multiple times of fault location, and take the mean value for final fault location result. Fig. 11 shows the distribution of 100 times of fault location results. The mean value is 24.36 km with the fault location error of 0.64 km (0.32%). In comparison, the fault location result using the full data group for training is 28.22 km with the error of 3.22 km (1.61%). This clearly proves the importance to adopt physics information to select the target datset during the data driven procedure.

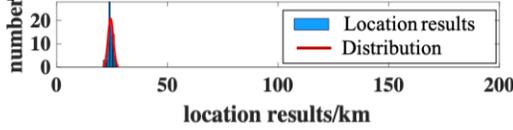

Fig. 11 Fault location results with multiple times, 1 ohm A-G fault at 25 km

### D. Fault Location for an Example High Resistance Fault

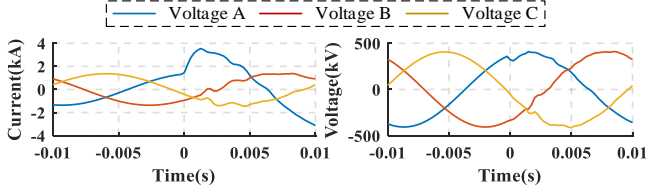

Fig. 12. Current and voltage measurements, 160 ohm A-G fault at 25 km

Take a 160 ohm A-G fault at 25 km as an example. The local terminal measurements are shown in Fig. 12. Due to space limitations, only the procedure to determine the fault resistance range and the fault location results are shown. Similar as Fig. 10, the range of fault resistance is determined as $R_f \in [10,190]\, ohm$, as shown in Fig. 13. One can observe that the estimated range correctly reflects that the fault is a high resistance fault. After 100 times of the fault location procedure, the distribution of results is shown in Fig. 14. The mean value of fault location is 24.02 km, with the error of 0.98 km (0.49%). In comparison, the fault location result using the full data group for training is 28.68 km with the error of 3.68 km (1.84%).

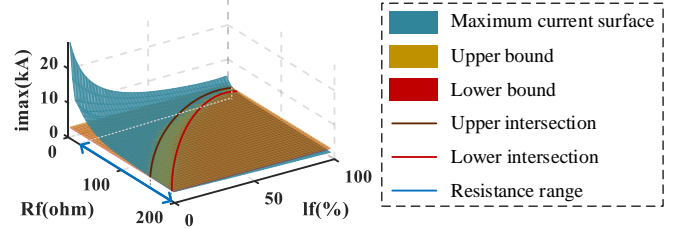

**Fig**. 13. The range of the fault resistance, 160 ohm A-G fault at 25 km

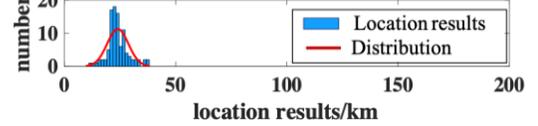

Fig. 14 Fault location results with multiple times, 160 ohm A-G fault at 25 km

### E. Fault Location for Various Fault Events

Various fault events with different fault types, locations and resistances are studied. The results are summarized in Fig.15 (a)-(e) and Table VII. The proposed method shows improved fault location accuracy than the traditional method without utilizing the physics information to select the target dataset.

The fault location method is implemented on a personal computer with Intel i7-7700 CPU. The parameter estimation procedure is implemented via Matlab, while the training procedure is implemented via python. In each fault event, the parameter estimation procedure takes less than 0.5 seconds. The training and testing procedure always takes less than 2 seconds for one fault location result. Considering the repetition of 100 times, the calculation burden of final fault location is always less than 4 minutes for each fault event. This calculation burden is acceptable in practice.

## VI. FIELD DATA EXPERIMENTS

This section will further test the proposed method with the field data. Here two field fault events on two 220 kV transmission lines are tested. For each transmission line system, the full data group is generated based on the system parameters shown in Table VIII, and the line parameters are selected from the utility database.

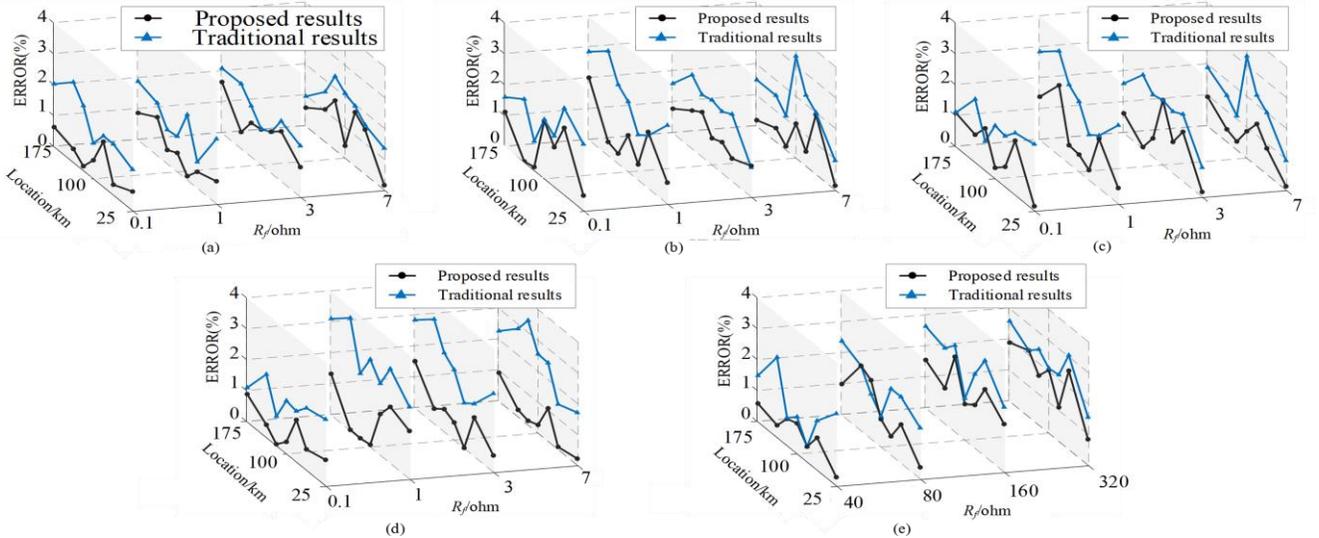

**Fig**. 15 Five different case (a) SLG fault with low resistance, (b) LL fault, (c) LLG fault, (d) three phase fault, (e) SLG fault with high resistance.

Table.VII Average error of the simulation results

| Single line to ground fault (AG-low) | | | | |
|---|---|---|---|---|
| Rf/ohm | 0.1 | 1 | 3 | 7 |
| Proposed | 0.61% | 0.65% | 1.0% | 1.1% |
| Traditional | 1.78% | 1.59% | 1.70% | 1.70% |
| Line to line fault (BC) | | | | |
| Rf/ohm | 0.1 | 1 | 3 | 7 |
| Proposed | 1.12% | 0.99% | 1.31% | 1.70% |
| Traditional | 1.66% | 2.39% | 1.88% | 1.80% |
| Line to line to ground fault (BCG) | | | | |
| Rf/ohm | 0.1 | 1 | 3 | 7 |
| Proposed | 0.87% | 1.03% | 0.88% | 0.97% |
| Traditional | 1.66% | 2.39% | 1.88% | 1.80% |
| Three phase fault (ABC) | | | | |
| Rf/ohm | 0.1 | 1 | 3 | 7 |
| Proposed | 0.74% | 0.88% | 0.73% | 0.41% |
| Traditional | 1.25% | 2.10% | 1.98% | 2.14% |
| Single line to ground fault (AG-high) | | | | |
| Rf/ohm | 40 | 80 | 160 | 320 |
| Proposed | 0.71% | 1.1% | 1.7% | 1.9% |
| Traditional | 1.50% | 1.84% | 2.22% | 2.15% |

Table.VIII Parameters in the full data group, field data

| Parameters | Value |
|---|---|
| Source impedance/ohm | $Z_\alpha = Z_\beta = 0.1+j1,\ 0.4+j1.5,\ 1+j5$; $Z_0 = 0.2+j1.5,\ 0.8+j3,\ 2+j6$ |
| Loading condition/deg | $\pm 2,\ \pm 6,\ \pm 10,\ \pm 12$ |
| Fault resistance/ohm | 0.01, 0.1, 0.2, 0.5, 1, 10, 20, 50, 100, 150, 200 |
| Fault location/km | 1, 2, 3, …, ⌊line length⌋ |
| FIA/deg | 0, 45, 90, …, 270, 315 (step: 45 deg) |

### A. Case 1: C-G fault at 11.9 km, 220 kV 22.60 km Line

Case 1 is a C-G fault at 11.9 km of a 220 kV, 22.60 km HVAC transmission line. The single-ended voltage and current waveforms recorded by the DFRs are stored in the COMTRADE file and are shown in Fig. 16. The sampling rate is 4 kHz. Table.IX shows the line parameters from the utility database.

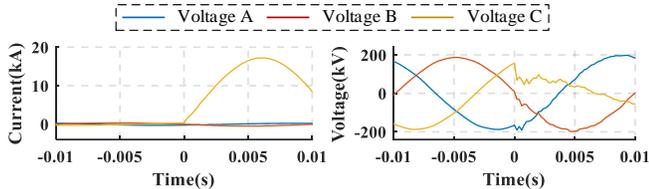

Fig.16 Current and voltage measurements, field data case1

Table.IX Line parameters, field data case 1

| $R$ (ohm/km) | $L$ (H/km) | $C$ (F/km) |
|---|---|---|
| $R_S = 0.0849$ | $L_S = 0.00141$ | $C_M = -9.134\text{e-}11$ |
| $R_M = 0.0449$ | $L_M = 4.47\text{e-}4$ | $C_S = 1.2166\text{e-}08$ |
| $R_0 = 0.1724$ | $L_0 = 0.00227$ | $C_0 = 1.225\text{e-}08$ |
| $R_1 = 0.0400$ | $L_1 = 9.548\text{e-}4$ | $C_1 = 1.198\text{e-}08$ |

*Step 1*: From section III, the following system parameters are estimated. From symmetry of the system, the phase rotation can be adopted to convert the C-G fault (with phase sequence A, B, C) into an equivalent A-G fault (with the phase sequence B, C, A). From (6), the *source impedance* is calculated as $Z_{s\alpha 1} = 1.2678+j6.7522$, $Z_{s01} = 1.5215+j11.3984$ ohm. From (7), the *loading condition* is calculated as 1.9916 deg. From (8), the *fault inception angle* is calculated as 57.6 deg. From (22), the calculation procedure of the *range of fault resistance* is illustrated in Fig. 17. From the current measurements, the value of $\max|i_{1,meas}^{\alpha}(t)|$ is 11.58 kA, and the upper bound/lower bound values are 12.16/11.00 kA. Similarly, the intersections between the "maximum current surface" and the upper/lower bound planes determine the range of fault resistance as $R_f \in [0, 3.3]\,ohm$.

*Step 2*: With the estimated system parameters in *Step 1*, the target dataset is selected within the full data group. The system parameters of the target dataset are shown in Table X.

*Step 3*: The target dataset is fed into the neural network for training. The parameters of the NN are the same as those in section V. Fig. 18 shows the distribution of 100 times of fault location results. The mean value of the proposed and the traditional methods are 12.003 km (error 0.103 km) and 16.334 km (4.434 km), respectively.

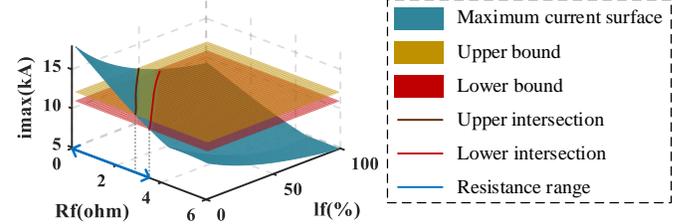

Fig. 17 The range of the fault resistance, field data case 1

Table.X Parameters in the target dataset, field data case 1

| Parameters | Value |
|---|---|
| Source impedance/ohm | $Z_\alpha = Z_\beta = 0.4+j1.5,\ 1+j5$; $Z_0 = 0.8+j3,\ 2+j6$ |
| Loading condition/deg | $\pm 2$ |
| Fault resistance/ohm | 0.01, 0.1, 0.2, 0.5, 1, 10 |
| Fault location/km | 1, 2, 3, …, 22 |
| FIA/deg | 45, 90 |

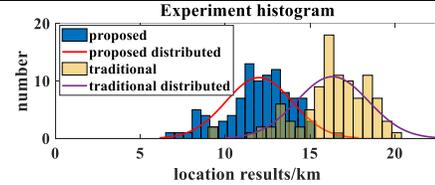

Fig. 18 Fault location results with multiple times, field data case 1

### B. Case 2: A-G fault at 1.8 km, 220 kV 23.55 km Line

Case 2 is a A-G fault at 1.8 km of a 220 kV 23.55 km HVAC transmission line. Similarly, the single-ended measurements from DFRs are shown in Fig. 19. The sampling rate is 5 kHz. Table.XI shows the line parameters from the utility database.

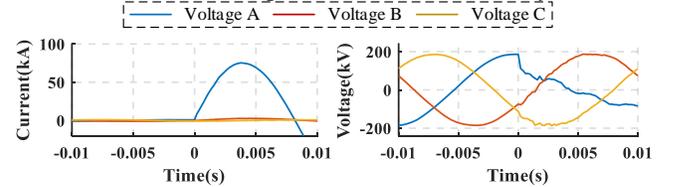

Fig.19 Current and voltage measurements, field data case 2

Table.XI Line parameters, field data case 2

| $R$ (ohm/km) | $L$ (H/km) | $C$ (F/km) |
|---|---|---|
| $R_S = 0.0734$ | $L_S = 0.0015$ | $C_M = -2.6673\text{e-}09$ |
| $R_M = 0.0367$ | $L_M = 4.6667\text{e-}04$ | $C_S = 1.1050\text{e-}08$ |
| $R_0 = 0.1469$ | $L_0 = 0.0024$ | $C_0 = 5.7156\text{e-}09$ |
| $R_1 = 0.0367$ | $L_1 = 0.0010$ | $C_1 = 1.3717\text{e-}08$ |

*Step 1*: From section III, the following system parameters can be similarly estimated. The *source impedance* is calculated as $Z_{s\alpha 1} = 0.2881+j0.8813$, $Z_{s01} = 0.1255+j1.0429$ ohm. The *loading condition* is calculated as 2.7417 deg. The *fault inception angle* is calculated as 97.2 deg. From the current measurements, the value of $\max|i_{1,meas}^{\alpha}(t)|$ is 49.38 kA, and the upper bound/lower bound values are 51.84/46.91 kA. The *range of fault resistance* is determined via Fig. 20, and is $R_f \in [0, 2.8]\,ohm$.

*Step 2*: With the estimated system parameters in *Step 1*, the target dataset is selected within the full data group. The system parameters of the target dataset are shown in Table XII.

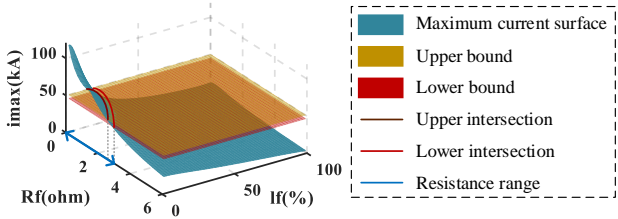

**Fig**. 20. The range of the fault resistance, field data case 2

**Table**. XII Parameters in the target dataset, field data case 2

| Parameters | Value |
| --- | --- |
| Source impedance/ohm | $Z_\alpha=Z_\beta=0.1+j1$, 0.4+j1.5; $Z_0=0.2+j1.5$, 0.8+j3 |
| Loading condition/deg | 6, 2 |
| Fault resistance/ohm | 0.01, 0.1, 0.2, 0.5, 1, 10 |
| Fault location/km | 1, 2, 3, …, 23 |
| FIA/deg | 45, 90 |

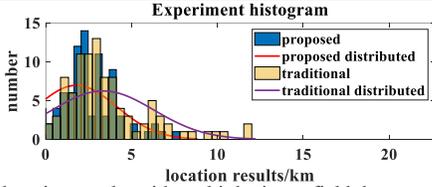

**Fig**. 21 Fault location results with multiple times, field data case 2

*Step 3*: The target dataset is fed into the neural network for training. The hyper parameters of the NN are the same as those in section V. Fig. 21 shows the distribution of 100 times of fault location results. The mean value of the proposed and the traditional methods are 1.777 km (error 0.023 km) and 3.337 km (1.5 km), respectively.

### C. Summary of Field Data Results

One can observe that field data will greatly affect the accuracy of the traditional method without dataset selection. This is because the mismatch between the practical power system and the simulation system will magnify the fault location error. On the contrary, with the physics information extracted from the field data, the selected target dataset can well represent the fault features of practical power systems, and therefore the proposed physics-informed data-driven method still presents high fault location accuracy.

For each fault event, the entire fault location procedure only takes about 2 minutes, including the parameter estimation (less than 0.5 sec) and the calculation of 100 fault location results (each fault location takes around 1.2 seconds). The algorithm implementation platform is the same as those in section V. This calculation burden is acceptable in practice.

## VII. DISCUSSION

This section makes further discussions about the proposed method. The two field data cases are taken as examlpes.

### A. Line Parameter Error

This section studies the influence of the line parameter errors on the proposed method. The 5% and 10% errors are added to all line parameters in the utility database. The fault location results for the field data case 1 and 2 are shown in Fig. 22 and 23, respectively. With parameter errors of 0%, 5%, and 10%, the fault location results are 12.003 km (0.103 km error), 11.263 km (0.637 km error) and 11.042 km (0.858 km error) for case 1, and 1.777 km (0.023 km error), 1.248 km (0.552 km error), and 1.243 km (0.557 km error) for case 2, respectively. The proposed method demonstrate some robustness towards parameter errors.

### B. Comparison to Model-Based Single-Ended Methods

In practice, model-based single-ended methods are widely adopted for fault location. Those methods are typically based on phasor representations, and therefore usually require a relatively long time window after the fault occurs for accurate extraction of phasors. However, for transmission lines protected by fast-tripping relays, the available time window during fault could be even less than 1 cycle, limiting the accuracy of phasor extraction and the phasor based fault location methods.

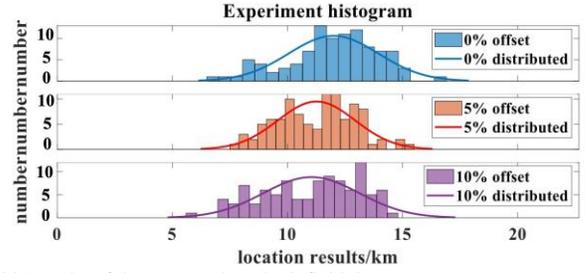

**Fig**. 22 Results of the proposed method, field data case 1, parameter error

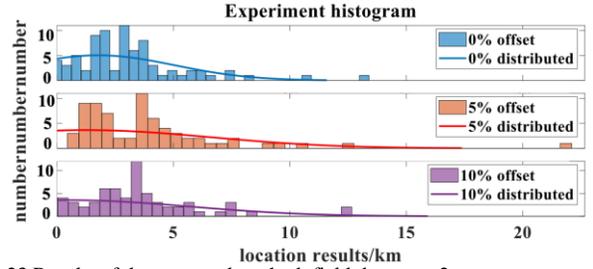

**Fig**. 23 Results of the proposed method, field data case 2, parameter error

Next, the widely adopted Takagi method [22] is shown as an example. For each time instant, the fundamental frequency phasors are extracted based on IEEE C37.118 synchrophasor standard [23], and the time window is selected within one cycle before the time instant.

Fig. 24 shows the fault location results with Takagi method for field data case 1. At 30 ms (around 1.5 cycles after the fault occurs), the system is close to steady state, and the fault location result is 12.6881 km (0.7881 km error). This fault location result is rather accurate. However, if the line is tripped fast, and only 10 ms data window is available (half a cycle after the fault occurs, as adopted by the proposed method), the Takagi method locates the fault at 31.9631 km (20.0631 km error). In comparison, the proposed method accurate locate the fault as 12.003 km (0.103 km error) using the fault data within half a cycle after the fault. This clearly proves the advantages of the proposed method.

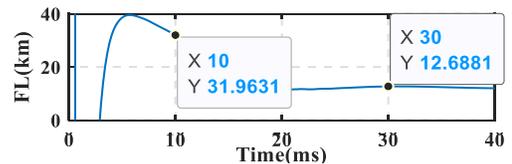

**Fig**. 24 Results of the Takagi method, field data case 1

Similarly, the fault location results for field data case 2 are shown in Fig. 25. Similarly, the fault location result is rather accurate at 1.5 cycles after the fault occurs (1.6819 km, with the error of 0.1181 km); however, the result at 10 ms (half a cycle) is 8.4379 km, with the large error of 6.6379 km. In comparison, the proposed method locates fault at 1.777 km (0.023 km error) only using the fault data within half a cycle after the fault.

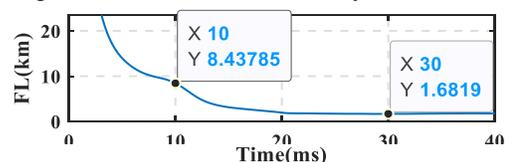

**Fig**. 25 Results of the Takagi method, field data case 2

## VIII. CONCLUSION

In this paper, a single-ended physics-informed data-driven method for transmission line fault location method is proposed. It only requires the single-ended measurements within half a cycle before and after the fault initiation time. With the fault data from a field fault event, the proposed method first extrat the range of system parameters, including equivalent source impedances, loading conditions, fault inception angles, and fault resistances. This procedure extracts physics information within the fault data. Second, with the estimated range of system parameters, the target dataset is generated using electromagnetic simulation tools, to embed the physics information into the target dataset. Through this procedure, the target dataset is consistent with the field data, with similar fault and system features. Finally, the target dataset is fed into the neural network for training for fault location. Simulation results confirms that the proposed method presents lower fault location error compared to the traditional data-driven methods without considering the physics information from the field data. Also, field data experiments further validate the accuracy of the proposed physics-informed data-driven method and prove the importance to consider physics information in practice. ompared to single-ended model based fault location methods, the proposed method does not need to extract phasors, and works with a short time window within half a cycle after the fault occurs. The proposed data-driven method is compatible with practical digital fault recorders with the relatively low sampling rate of several kilo samples per second, and the computational burden is acceptable for practical implementation.

## APPENDIX. EQUIVALENCE OF TWO BRANCHES IN PARALLEL

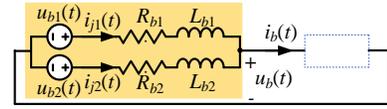

Fig. A1. Two Thevenin's branch in parallel

This section derives the time domain equivalencing of two Thevenin's branch in parallel, as shown in Fig. A1.

$$0 = -u_b(t) + u_{b1}(t) + R_{b1}i_{b1}(t) + L_{b1}\,di_{b1}(t)/dt$$
$$0 = -u_b(t) + u_{b2}(t) + R_{b2}i_{b2}(t) + L_{b2}\,di_{b2}(t)/dt \quad (A-1)$$
$$0 = -i_b(t) + i_{b1}(t) + i_{b2}(t)$$

Transform the time domain equaiton (A-1) into Laplace domain (s domain),

$$0 = -u_b(s) + u_{b1}(s) + R_{b1}i_{b1}(s) + sL_{b1}i_{b1}(s)$$
$$0 = -u_b(s) + u_{b2}(s) + R_{b2}i_{b2}(s) + sL_{b2}i_{b2}(s) \quad (A-2)$$
$$0 = -i_b(s) + i_{b1}(s) + i_{b2}(s)$$

The simplification is achieved by describing the relationship between $u_b(s)$ and $i_b(s)$. In other words, eliminating $i_{b1}(s)$ and $i_{b2}(s)$ in (A-2). This is achieved with the approximation that $R_{b1}/R_{b2} \approx L_{b1}/L_{b2}$. For the cases within this paper, the resistance and the inductance are obtained by the sum of line and source line parameters. The above approximation is reasonable. With the approximation, the simplified relationship is,

$$u_b(s) - \frac{R_{b2}u_{b1}(s) + R_{b1}u_{b2}(s)}{R_{b1} + R_{b2}} = \left[\frac{R_{b1}R_{b2}}{R_{b1} + R_{b2}} + s\frac{L_{b1}L_{b2}}{L_{b1} + L_{b2}}\right]i_b(s) \quad (A-3)$$

Transform Laplace domain equation (A-3) into time domain equation,

$$u_{seq}(t) - u_b(t) = R_{eq}i_b(t) + L_{eq}\,di_b(t)/dt \quad (A-4)$$

where $u_{seq}(t) = [R_{b2}u_{b1}(t) + R_{b1}u_{b2}(t)]/[R_{b1} + R_{b2}]$, $R_{eq} = R_{b1}R_{b2}/[R_{b1} + R_{b2}]$, and $L_{eq} = L_{b1}L_{b2}/[L_{b1} + L_{b2}]$. This equation corresponds to the circuit diagram shown in Fig. A2.

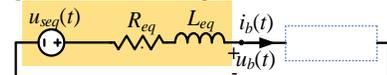

Fig. A2. Equivalence of two Thevenin's branch in parallel